\documentclass[twocolumn,showpacs,preprintnumbers,amsmath,amssymb,prb,superscriptaddress,aps,10pt]{revtex4-1}

\usepackage{graphicx}
\usepackage{bm}

\newcommand{\cumu}[1]{\langle\!\langle #1 \rangle\!\rangle}

\begin{document}

\title{Current noise and higher order fluctuations in semiconducting bilayer systems}
\author{H. Soller}
\affiliation{Institut f\"ur Theoretische Physik,
Ruprecht-Karls-Universit\"at Heidelberg,\\
 Philosophenweg 19, D-69120 Heidelberg, Germany}
\author{A. Komnik}
\affiliation{Institut f\"ur Theoretische Physik,
Ruprecht-Karls-Universit\"at Heidelberg,\\
 Philosophenweg 19, D-69120 Heidelberg, Germany}
\date{\today}

\begin{abstract}
We analyze the transport properties of a semiconductor based bilayer system under non-equilibrium conditions with special emphasis on the charge transfer statistics in the regime dominated by the exciton transport. We consider two different models. In one of them the transport occurs incoherently and is dominated by incoherent tunneling processes of individual excitons, while in the other system no disorder is present and transport processes are fully coherent.  
We find that the strength of cross correlations of currents in different layers is only insignificantly affected by the disorder and shows up similar behaviour in both systems. We discuss possible experimental realizations and make predictions for measurable quantities. 
\end{abstract}


\maketitle

\section{Introduction} \label{s1}

Electronic bilayer systems have recently drawn increasing attention due to new experimental results on electronic transport \cite{2012arXiv1206.6626G,2012arXiv1203.3208N,su2008} as well as due to the discovery of the novel macroscopic ordering driven by quantum effects.\cite{eisenstein1,gossard,1367-2630-14-6-063010,2012arXiv1211.6562P} In a generic situation such systems consist of two spacially separated and mutually insulated semiconducting layers which can be individually gated. On the other hand, the interlayer separation is engineered in such a way that the Coulomb interaction between majority carriers in both subsystems is still large. This can be achieved by an appropriate choice for the barrier material. \cite{PhysRevB.78.121401,2011LTP....37..583B} In such a situation the formation of excitons is possible.  Even a condensation of such electron-hole pairs was predicted \cite{snoke,lozovik,PhysRev.126.1691,2012NRL.....7..145P} and subsequently observed in several experiments performed on GaAs quantum wells using AlGaAs as the insulating barrier at total filling factor $\nu = 1$ in the quantum Hall regime. \cite{PhysRevLett.68.1383,PhysRevLett.80.1714,PhysRevLett.84.5808,PhysRevLett.93.036801,PhysRevLett.93.036802,1367-2630-10-4-045018,PhysRevLett.106.236807} 


While the non-linear transport characteristics of such systems have been investigated before \cite{su2008}, their noise properties received only little attention. In \onlinecite{PhysRevLett.108.156401} we tried to close this gap and have calculated the cumulant generating function (CGF) of a bilayer system attached to four different electrodes. Some very interesting effects were found, which can only be attributed to the non-trivial interactions between the layers. In order to perform the calculations one had to assume that the transport processes at different junctions occur in an incoherent fashion (sequential tunneling). Despite this severe restriction the currents in both layers are correlated due to the locality of the bound electron-hole pairs. Here we would like to drop this assumption and consider a fully coherent system. 

The paper is organized as follows. In the next section we introduce the system and give the details of its mathematical implementation. We briefly review the main results of \onlinecite{PhysRevLett.108.156401} and adapt them to a recent experimental setup presented in \onlinecite{2012arXiv1203.3208N}. 
In Section \ref{s3} we present a toy model consisting of two short Hubbard chains, which are coupled with each other, so that a formation of an interchain electron-hole pair or exciton is adequately described. In the next step we calculate the full counting statistics, compare the predictions of both models, analyze their similarities and discrepancies. We briefly summarize the results in Section \ref{s4}.

%
%
%
%

\section{Incoherent tunneling approximation} \label{s2}

\begin{figure}[th]
\centerline{\includegraphics[width=7cm]{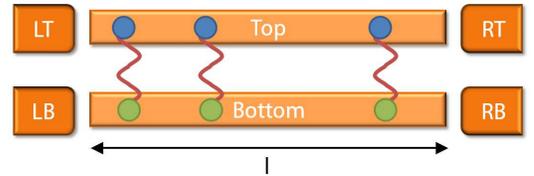}}
\vspace*{8pt}
\caption{Sketch of the experimental setup. A semiconducting bilayer is contacted by four metallic leads denoted by the indices LT (left top), LB (left bottom), RT and RB.}
\label{fig1}
\end{figure}

The basic system we study is depicted in Fig.~\ref{fig1}: a semiconducting bilayer is contacted by four metallic electrodes via tunnel contacts. The Hamiltonian of the system reads \cite{PhysRevLett.108.156401}
\begin{eqnarray}
H = H_N + H_T + H_{\mathrm{SC}}, \label{htotal}
\end{eqnarray}
where the term $H_N$ describes the four metallic electrodes which are described as fermionic continua written in terms of electron field operators $\alpha_\sigma$ at chemical potentials $\mu_{\alpha\sigma}$. $\alpha = L,R$ refers to the contacts on the left/right side of the respective layer and $\sigma = T,B$ labels the top and bottom layer, respectively. We assume the leads to be in the wide flat band limit so that their density of states (DOS) $\rho_0$ is constant and denote the Fermi distribution functions in the respective electrode by $n_{\alpha\sigma}$. 
%
%
For the phenomena we would like to consider the spin degree of freedom is irrelevant, 
therefore we work with a spinless system.

$H_T$ is the tunneling Hamiltonian describing hopping between each layer of the bilayer and the corresponding metallic electrodes
\begin{eqnarray}    \label{orHT}
H_T = \sum_{\substack{\sigma = T,B \\ \alpha = L,R}} \gamma_{\alpha \sigma} (\alpha_\sigma^+ \Psi_\sigma + \Psi_\sigma^+ \alpha_\sigma), \label{htunnel}
\end{eqnarray}
using tunneling amplitudes $\gamma_{L, T/B}, \; \gamma_{R, T/B}$. Finally, $H_{\mathrm{SC}}$ describes the bilayer using a simple one-dimensional model as in \onlinecite{PhysRevLett.108.156401,PhysRevLett.104.027004} (we shall discuss generalizations to higher dimensions later)
\begin{eqnarray}      \label{HSC}
H_{\mathrm{SC}}= \int_{-l/2}^{l/2} \!\!\! dx  \, \,   \Psi^\dagger(x)  \left(\begin{array}{cc} H_{T} & \Delta  \\ \Delta^*  & H_{B} \end{array}\right) \Psi^{}(x) \label{hec} \, ,
\end{eqnarray}
where $l$ is the longitudinal distance between the right and left side of the semiconductor, $\Psi = (\Psi_T, \Psi_B)^T$ is the two-layer spinor and $H_T, H_B$ describe the electron/hole single particle Hamiltonian of the top and bottom layer. We describe the interlayer Coulomb interaction by an exciton order parameter $\Delta(x)$, which in general is a space-dependent quantity. \cite{PhysRevB.78.121401} Its absolute value $\Delta_0$ in equilibrium represents the excitonic gap and directly gives the exciton coherence length $\xi_{\mathrm{SC}} = v_F / \Delta_0$. It can be determined self-consistently from the interlayer interaction as in BCS theory.\cite{comte}\\

In order to calculate the cumulant generating function (CGF) of charge transfer we use a Green's function (GF) method developed for quantum impurities in \onlinecite{PhysRevB.73.195301}. For every electrode present in the system we have to introduce a fictitious measuring field $\mbox{\boldmath$\lambda$} = (\lambda_{LT}, \lambda_{LB}, \lambda_{RT}, \lambda_{RB})$ which counts an electron when it crosses the electrode-semiconductor junction in question. In this manner fluctuations, either caused by the discreteness of charge or thermally induced, can be taken care of. The measuring fields are both contour and time-dependent: they are nonzero only during the measurement time $\tau$ and change sign on the $+$ and $-$ branch of the Keldysh contour. This procedure allows for the calculation of the CGF for arbitrary parameters in Eq. (\ref{htotal}) as discussed in \onlinecite{PhysRevB.54.7366} and has already been applied to numerous quantum impurity problems (see e.g. \onlinecite{PhysRevB.84.235408,PhysRevB.80.041309,PhysRevB.84.155305}).

The CGF can be calculated as a generalized Keldysh partition function
\begin{eqnarray}
\chi(\mathbf{\lambda}) = \langle T_{\cal C} e^{-i \int_{\cal C} T^{\mathbf{\lambda}}(t) dt}\rangle,
\end{eqnarray}
where $T_{\cal C}$ is the Keldysh contour ordering operator and $T^{\mathbf{\lambda}}(t)$ can be derived from the tunnel Hamiltonian in Eq. (\ref{htunnel}) using the substitution $\alpha_\sigma \rightarrow \alpha_\sigma e^{i \lambda_{\alpha\sigma}(t)/2}$. In order to calculate the CGF we then rewrite the expression using the adiabatic potential method in the limit of long measurement times $\tau$ as $\ln \chi(\mathbf{\lambda}) = - i \tau U(\mathbf{\lambda})$. The adiabatic potential $U(\mathbf{\lambda})$ may in turn be rewritten in terms of the bare and exact-in-tunneling GFs of the system described by Eq.~(\ref{htotal}).

The Hamiltonian in Eq.~(\ref{hec}) is very similar to the BCS Hamiltonian. Indeed, $T$ and $B$ can be interpreted as pseudospin indices to map this Hamiltonian with the additional exchange of $B$ holes to electrons onto BCS theory.\cite{lozovik2} Also one has to take into account that electrons and holes each have their own chemical potential $\mu_{\mathrm{SC},\sigma}, \; \sigma = T, B$.\cite{comte} Consequently we can also rephrase our problem as the calculation of the spin-dependent CGF for a normal-superconductor-normal (NSN) junction with different chemical potentials for the spin species. We want to stress that despite the modelling of the condensate as a perfect homogeneous system in describing the tunneling we assume for the dephasing time $\tau_\phi$: $\xi_E \ll v_F \tau_{\phi} \ll l$ (incoherent tunneling approximation).

The CGFs for different normal-superconductor hybrid structures have been considered before. \cite{PhysRevLett.87.067006,PhysRevB.50.3982,PhysRevB.79.054505,springerlink:10.1140/epjd/e2010-00256-7,soller1} Here the task is to work out the spin-dependent case, to take care of the presence of electrons instead of holes in the lower layer and to address the case of a bilayer contacted from two sides. However, the $\Delta(x)$-coupling in Eq.~\eqref{HSC} is formally a tunneling term. Since the layers are supposed to be isolated from each other we have to suppress particle current flowing between the layers. This is very conveniently done by adjusting the chemical potentials in the layers self-consistently. 
A similar procedure was implemented for a normal-superconductor-normal junction in Ref.~\onlinecite{lambert}. Under such circumstances it is then unimportant whether we count charges on the left side or the right side of the respective bilayer. This means that we can then calculate the CGF only for the left side and obtain the one for the right side by exchanging $\lambda_{LT} = - \lambda_{RT}, \; \lambda_{LB} = - \lambda_{RB}$. We would like to point out that this self-consistency condition is very different from the conventional one inspired by the BCS theory as it defines the  \emph{relative}  position of the exciton chemical potential with respect to the ones in the external electrodes. 

Using the microscopic Hamiltonian approach as outlined in Ref. \onlinecite{PhysRevB.54.7366} we have obtained the complete expression for the temperature and energy dependent CGF, valid for arbitrary values of the model parameters. \cite{PhysRevLett.108.156401} The expression, which has been used for our numerical calculations, is quite lengthy and we do not report it here. Nevertheless, all relevant ingredients of the CGF already appear in the limits of small bias ($\mu_{L\sigma} - \mu_{R\sigma} \ll \Delta, \; \sigma=T,B$) and large bias  ($\mu_{L\sigma} - \mu_{R\sigma}  \gg \Delta, \; X=T,B$), where the expression of the CGF greatly simplifies. Indeed in these regimes the non-interacting self-energy due to the EC is either real and purely off-diagonal or imaginary and purely diagonal, like in a superconductor \cite{PhysRevLett.80.2913}, while the one due to the normal lead is always diagonal. In these regimes the CGF acquires the following expression
\begin{widetext}
\begin{eqnarray}
&& \ln \chi_L(\lambda_{LT}, \lambda_{LB}) = 4 \tau \int \frac{d\omega}{2\pi} \left[ \sum_{\sigma = T,B} \ln \left\{1+ T_\sigma(\omega) \left[(e^{i \lambda_{L\sigma}} -1) n_{L\sigma} (1-n_{\mathrm{SC},\sigma}) \right. \right. \right. \nonumber\\
&& \left. \left. + (e^{-i \lambda_{L\sigma}} -1) n_{\mathrm{SC},\sigma} (1-n_{L\sigma})\right]\right\} \theta\left(\frac{|\omega_\sigma| - \Delta}{\Delta}\right) \nonumber\\
&& + \ln \left\{1+ T_A(\omega) \left[(e^{i \lambda_{LT}} e^{-i \lambda_{LB}} -1) n_{LT} (1-n_{LB}) \right. \right. \nonumber\\
&& \left. \left.\left. + (e^{i \lambda_{LB}} e^{-i \lambda_{LT}} -1) n_{LB} (1-n_{LT}) \right]\right\}\theta\left(\frac{\Delta - \max(|\omega_T|, \; |\omega_B|)}{\Delta}\right)\right], \label{cgf}
\end{eqnarray}
\end{widetext}
where $\omega_{T,B} = \omega - \mu_{\mathrm{SC},\sigma}, \; \sigma=T,B$ and the transmission coefficients are given by
\begin{eqnarray}
T_\sigma(\omega) = \frac{4\tilde{\Gamma}_{L\sigma}}{(1+ \tilde{\Gamma_{L\sigma}})^2} \;\; \mbox{and} \;\; T_A(\omega) = \frac{4\tilde{\Gamma}_A}{(1+ \tilde{\Gamma}_A)^2}.
\end{eqnarray}
The energy-dependent DOS of the EC affects the hybridisations
\begin{eqnarray*}
\tilde{\Gamma}_{L\sigma} &=& \frac{\Gamma_{L\sigma} |\omega_\sigma|}{\sqrt{\omega_\sigma^2 - \Delta^2}}, \;\; \mbox{and} \;\; \tilde{\Gamma}_A = \frac{\Gamma_{LT} \Gamma_{LB} \Delta^2}{\sqrt{\Delta^2 - \omega_T^2} \sqrt{\Delta^2 - \omega_B^2}},
\end{eqnarray*}
where $\Gamma_{L\sigma} = \pi^2 \rho_{0L\sigma} \rho_{0\mathrm{SC}} \gamma_{L\sigma}^2/2$, with the DOS of the EC given by $\rho_{0\mathrm{SC}}$. $n_{\mathrm{SC}\sigma}$ refers to a Fermi distribution function at chemical potentials $\mu_{\mathrm{SC},\sigma}, \; \sigma=T,B$ characterising the layer of the semiconductor. We use units such that $e = \hbar = k_B = 1$ and $G_0 = 2e^2/h$.

The first line of the CGF describes the supra-gap contribution, which is only due to single electron transport in the simplified form given here and characterised by the normal transmission coefficient $T_{\sigma}(\omega)$. The second line describes the sub-gap contribution due to excitonic Andreev reflection \cite{PhysRevLett.104.027004} in which an electron and a hole (in different layers) enter or leave the EC. $\mu_{\mathrm{SC},\sigma}, \; \sigma=T,B$ now have to be chosen such that \cite{lambert} 
\begin{eqnarray}
\cumu{I_{LT}} = - \frac{i}{\tau} \frac{\partial}{\partial \lambda_{T}} \chi_L = \cumu{I_{RT}}, \; \cumu{I_{LB}} = \cumu{I_{RB}}. \label{self}
\end{eqnarray}
For the sake of simplicity we consider the symmetric case $\Gamma_{LT} = \Gamma_{RT}$ and $\Gamma_{LB} = \Gamma_{RB}$. Then the requirement Eq.~(\ref{self}) is always fulfilled for  $\mu_{\mathrm{SC},T} = \mu_{\mathrm{SC},B} = 0$ if $\mu_{LT} = - \mu_{RT} = - V_T/2$ and $\mu_{LB} = - \mu_{RB} = -V_B/2$. This condition also implies that no current is flowing between the upper and lower layer.\cite{PhysRevB.56.3296}

 It is interesting to use the above formula to test fluctuation theorems. We have done so for the full result but for the sake of simplicity we illustrate the test for Eq. (\ref{cgf}). One of the most celebrated examples is the Cohen-Gallavotti relation, which can be directly related to a symmetry of the CGF.\cite{PhysRevLett.71.2401,PhysRevLett.74.2694,PhysRevLett.107.100601} For a single counting field $\lambda$ the CGF is expected to obey in the limit of long measurement time \cite{RevModPhys.81.1665}
\begin{eqnarray}
\ln \chi (\lambda) = \ln \chi (-\lambda - i \beta V) \, , \label{simplecohen}
\end{eqnarray}
where $\beta$ is the inverse temperature. 
However, in our case we encounter several complications compared to this simple case. First, the presence of an interlayer Coulomb interaction gives rise to the subgap contribution in Eq. (\ref{cgf}) which couples the top and bottom layer and therefore requires to extend the simple relation in Eq. (\ref{simplecohen}) to both $T$ and $B$ counting fields. Additionally different voltages can be applied on the top and bottom layer. Finally, we had to mind self-consistency which we satisfy by setting $\mu_{\mathrm{SC}, T} = \mu_{\mathrm{SC}, B} = 0$. Still, a relation similar to Eq. (\ref{simplecohen}) holds, namely \cite{PhysRevB.78.115429}
\begin{eqnarray*}
\ln \chi (\lambda_{LT}, \lambda_{LB}) = \ln \chi (-\lambda_{LT} - i \beta V_T/2, -\lambda_{LB} - i \beta V_B/2).
\end{eqnarray*}
Even though this is a straightforward generalisation of the Cohen-Gallavotti relation in Eq. (\ref{simplecohen}) the above relation provides an example how correlations affect the fluctuations in quantum systems. In this way we have supplied an example of a system where a generalized Cohen-Gallavotti fluctuation theorem holds.


From the CGF in (\ref{cgf}) we can read off several interesting transport features. At $T=0$ and $|V_T|, |V_B| < 2\Delta$ the excitonic Andreev reflections dominate the charge transport. Then we immediately obtain
\begin{eqnarray}
\cumu{{I_{LT}}} = - \frac{i}{\tau} \frac{\partial}{\partial \lambda_{LT}} \chi_L = \frac{i}{\tau} \frac{\partial}{\partial \lambda_{LB}} \chi_L = - \cumu{I_{LB}}. \label{counter}
\end{eqnarray}
Consequently, we correctly predict that the excitonic Andreev reflections induce counterpropagating currents in the layers.\cite{su2008} Depending on the bias choice we observe different transport features. For equal bias in the two layers $V_B =V_T$ the magnitude of the current becomes zero (exciton blockade). If only one layer is biased an opposite current in the other layer is induced (drag). If we further increase the bias difference between the layers the counterflowing currents increase.\cite{PhysRevLett.108.156401} This resembles the transport features observable under phase bias instead of voltage bias.\cite{PhysRevLett.104.027004,PhysRevB.84.184528}

\begin{figure*}[th]
\centerline{\includegraphics[width=13cm]{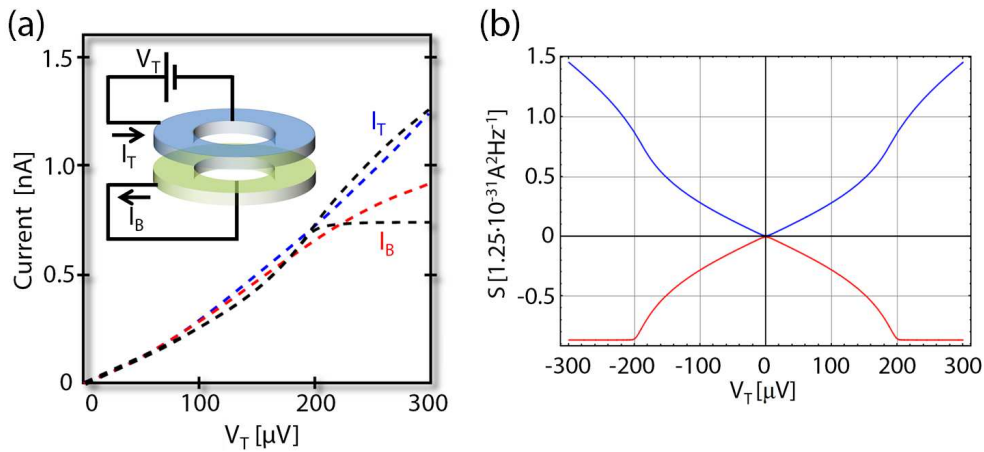}}
\vspace*{8pt}
\caption{(a): comparison of the prediction for the current using Eq. (\ref{cgf}) for a typical drag experiment in ECs. We use fitting parameters $\Gamma_T = 0.02$, $\Gamma_B = 0.72$, $\Delta = 100 \, \mu$eV and take the temperature $T=17$mK from experiment. The dashed blue and red curve correspond to the top and bottom current data from the experiment, respectively and the dashed black curves are the results from our model.\newline
(b): noise for the top current (upper blue curve) and the cross correlation of the top and bottom current (lower red curve) using Eq. (\ref{cgf}) and the same fitting parameters as in (a) for the  experiment \onlinecite{2012arXiv1203.3208N}.}
\label{fig2}
\end{figure*}

So far only experiments in a typical drag configuration have been made. The inset of Fig. \ref{fig2} (a) shows a sketch of the experiment described in \onlinecite{2012arXiv1203.3208N}, where two Corbino disks are realized in a GaAs/AlGaAs double quantum well structure. The experiment is performed in the $\nu =1$ quantum Hall state of the system and we expect exciton formation between the two Corbino disks that serve as the top and bottom layer indicated in Fig. \ref{fig1}. In the top layer a current can be excited and additionally the induced drag current in the bottom layer is recorded.\\

We calculate the current from Eq. (\ref{cgf}) and use $\Gamma_T = 0.02$, $\Gamma_B = 0.72$, $\Delta = 100 \; \mu$eV. We observe good qualitative agreement in Fig. \ref{fig2} (a).
\\
We only observe a slight deviation in the bottom current for voltages above the exciton gap. We ascribe this discrepancy to a small interlayer tunnel coupling that might still be present in the system and is also suggested from the analysis of the experimental data \onlinecite{2012arXiv1203.3208N}.


The next logical step is to compute the higher cumulants of the FCS. A number of details was already given in Ref.~\onlinecite{PhysRevLett.108.156401}. 
The current cross correlation of the top and bottom current for the setup shown in Fig.~\ref{fig2} (a) is plotted in Fig.~\ref{fig2} (b) using the same parameters. 
As we see from Fig. \ref{fig2} (b) one does not observe a positive cross correlation. This is very different from the setup in which a superconductor is contacted to two normal drains, in which one observes a positive cross correlation of the two currents in the normal leads via crossed Andreev reflection. \cite{PhysRevB.63.165314,2009arXiv0910.5558W,springerlink:10.1140/epjd/e2010-00256-7}  This discrepancy is due to the fact that we observe the correlation of electrons and holes and not correlations of electron pairs as in the case of superconductors. In this case the positive cross-correlation mediated by crossed Andreev reflection turns into an anti-correlation. We expect that this negative correlations should be observable in the experiments
similar to the one presented in Ref.~\onlinecite{2012arXiv1203.3208N}

Up to now we have considered a 1D system. Due to pointlike contacts to bias electrodes all excitations travel as $s$-waves even in higher-dimensional systems, in particular in genuine 2D layers. On the other hand, the ground state changes due to $\Delta(x)$ coupling between the layers are qualitatively the same in any dimension since the diagonalization procedure is dimension-independent. In connection with the assumption of incoherent tunnelling we conclude that the physical picture obtained above also holds for 2D layered systems. We expect fundamental changes to occur in the case of extended contacts when a number of different transport channels are possible. However, as long as these channels are non-interacting the corresponding FCS would be a simple sum of expressions for individual channels. A very similar procedure was considered in \onlinecite{PhysRevB.69.140502} in connection with superconducting contacts.


\section{Coherent transport: a toy model}
\label{s3}

\begin{figure}[th]
\centerline{\includegraphics[width=6cm]{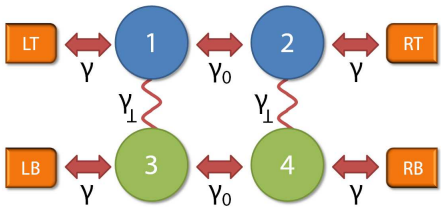}}
\vspace*{8pt}
\caption{Sketch of the toy model for coherent transport. The semiconducting bilayer in Fig. \ref{fig1} has been replaced by two pairs of tight binding chains which have an excitonic coupling described by $\gamma_{\bot}$.}
\label{fig4}
\end{figure}

\begin{figure*}[th]
\centerline{\includegraphics[width=12cm]{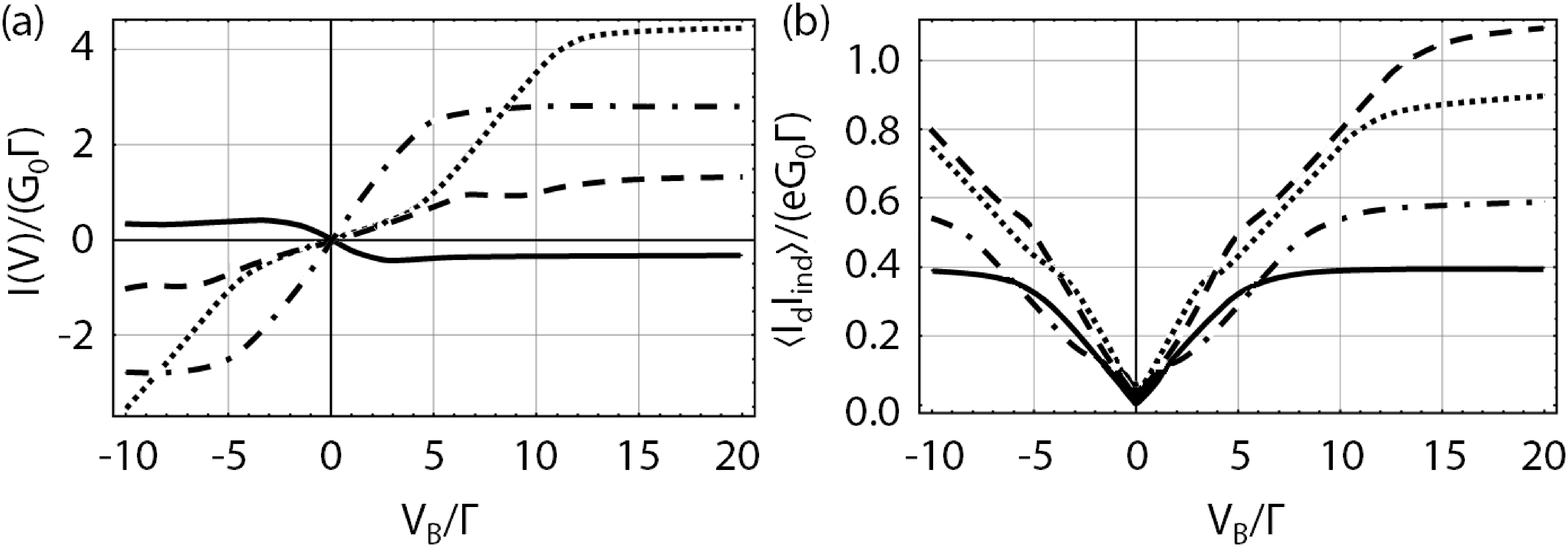}}
\vspace*{8pt}
\caption{Plots of the current and the cross correlation of currents. (a) shows the current $I$ as a function of voltage $V$ for $\gamma_0 = \Gamma$ and $T=0.05 \Gamma$. The solid black curve corresponds to the induced current for an excitonic coupling $\gamma_{\bot} = 0.85 \Gamma$ while the direct current is shown as the dotted dashed curve. The dashed black curve refers to the induced current for $\gamma_{\bot} = 4 \Gamma$ and the direct curve is shown as the dotted curve.\newline
In (b) we show the cross correlation of the direct and induced current for the same temperature and $\gamma_0$. The solid black curve is for $\gamma_{\bot} = 0.85 \Gamma$, the dotted dashed black curve shows the result for $\gamma_{\bot} = 2 \Gamma$, the dotted curve refers to $\gamma_{\bot} = 3 \Gamma$ and the dashed black curve represents $\gamma_{\bot} = 4 \Gamma$.}
\label{fig5}
\end{figure*}

Now we would like to abandon restrictions we made in the previous section. An appropriate model would be two non-interacting fermionic continua particle-hole coupled and perfectly clean so that $\tau_\phi \to \infty$ can be safely assumed. Such structure is, however, quite difficult to handle. Therefore we model the layers by tight-binding chains of finite lengths. It turns out that in order to account for most of the transport physics we only need to keep two sites in each chain. The effective model can then schematically represented as shown in Fig.~\ref{fig4}.
\\
The pairs 1,2 and 3,4 represent different layers of the device. The chemical potentials are now replaced by the bare energies of the sites $\mu_i$. Then the bilayer part of the Hamiltonian is given by
\begin{eqnarray}
 H_{B} &=& \sum_{k=1}^4 \mu_k \, d^\dag_k \, d_k + \left[ \gamma_0 (d^\dag_1 \, d_2 + d^\dag_3 \, d_4 ) \right. \nonumber\\
 && \left. +  \gamma_\perp (d^\dag_1 \, d_3 + d^\dag_2 \, d_4 ) + \mbox{H.c.}
 \right] \, ,
\end{eqnarray}
where $\gamma_0$ is the tunneling amplitude between the sites within the respective chain and $\gamma_\perp$ is the exciton coupling. $\gamma_0$ corresponds to the Fermi velocity of a continuum realization whereas $\gamma_\perp$ in the equivalent of the exciton coupling $\Delta(x)$ (it is measured in different units though). Coupling to the electrodes is modeled in a way similar to that of Eq.~\eqref{orHT}, 
\begin{eqnarray}
H_T = \gamma ( L_T^+ d_1 + L_B^+ d_3 + R^\dag_T d_2 + R^\dag_B d_4 + \mbox{H.c.} ) \, .
\end{eqnarray}
For simplicity we have chosen all tunneling amplitudes to the electrodes to be of the same magnitude and the DOS in each electrode to be $\rho_0$. In order to compute the FCS one performs the same transformation as shown in the previous sections in order to generate $T^\lambda$. After that the cumulant generating function can be evaluated using functional integration. We integrate the electrode degrees of freedom first, thereby obtaining the new effective Green's functions for the individual sites of the bilayer, 
\begin{eqnarray*}
  {\bf D}_k^{-1} = \left(
 \begin{array}{cc}
  \omega - \mu_k - i \Gamma (2 n_k - 1) &
  2 i e^{i \lambda_k} \, \Gamma n_k  \\
  - 2 i e^{- i \lambda_k} \, \Gamma ( 1 - n_k )  & - \omega + \mu_k - i \Gamma ( 2 n_k - 1) \end{array} \right) \, ,
\end{eqnarray*}
where $k=1,\dots, 4=LT,RT,LB,RB$, $\Gamma = \pi \rho_0 \gamma^2$ and $\lambda_k$ are the respective counting fields. Then the CGF is written as 
\begin{eqnarray}
 \chi({\boldsymbol{\lambda}}) = \int {\cal D}[{\bf d}] \, e^{-S} \, ,
\end{eqnarray}
where 
\begin{eqnarray}
  S =  {\bf d}^{\dag} \, {\bf K} \,  {\bf d} \, , {\bf K} = \left(
\begin{array}{cccc}
 D_1^{-1} & \gamma_{0} \sigma_z &  \gamma_{\perp} \sigma_z & 0 \\
  \gamma^*_{0} \sigma_z & D_2^{-1} & 0 &  \gamma_{\perp} \sigma_z  \\
 \gamma^*_{\perp} \sigma_z & 0 &  D_3^{-1} &  \gamma_{0} \sigma_z \\
 0 &  \gamma^*_{\perp} \sigma_z &  \gamma^*_{0} \sigma_z & D_4^{-1}
\end{array}
 \right) 
\end{eqnarray}
${\bf d}$ are superfields defined by
\begin{eqnarray}
  {\bf d} = \left( 
 d_{1-}, d_{1+}, d_{2-}, d_{2+}, d_{3-}, d_{3+}, d_{4-}, d_{4+} 
 \right) \, .
\end{eqnarray}
Here the index $\pm$ denotes the Keldysh component of the respective field. After the calculation of the functional integral we finally obtain
\begin{eqnarray}
 \ln  \chi({\boldsymbol{\lambda}}) = {\cal T} G_0 \int d \omega \ln \left[ \mbox{det} {\bf K}({\boldsymbol{\lambda}})/ \mbox{det} {\bf K}({\boldsymbol{\lambda=0}}) \right] \, .
\end{eqnarray}
$\mu_k$ are yet to be appropriately chosen. To fix them we perform the same procedure as in Section \ref{s2}. We compute the currents on both junctions of the upper layer and equalize them. On the other hand, from symmetry arguments it turns out that the choice $\mu_T=\mu_B=0$ is particularly convenient. One interesting parameter constellation is the situation of no bias in one of the layers (the upper one for definiteness) and finite voltage applied symmetrically around $\mu_B$ in the bottom subsystem.  One can produce analytical formulas for the direct current in the bottom layer ($\nu=d$) as well as for the induced current in the top layer ($\nu={\rm ind}$), valid at all temperatures,  
\begin{eqnarray}
 I_\nu(V)= \frac{G_0}{2} \int d \omega D_\nu(\omega) [ n_{LB}(\omega) - n_{RB}(\omega)] \, ,
\end{eqnarray}
where $n_{\alpha}(\omega)$ are the Fermi distribution functions in the respective electrodes. 
The effective transmission coefficients $D_\nu(\omega)$ have an interesting structure. They are particularly simple under the assumption  $\gamma_0=\gamma_\perp=\gamma/2$,
\begin{eqnarray*}
 D_d(\omega) &=& \frac{ \gamma^2 \Gamma^2 [\gamma^2 + 3 (\Gamma^2 + \omega^2)]}{(\Gamma^2 + \omega^2) [\gamma^4 + 2\gamma^2 (\Gamma^2 - \omega^2) + (\Gamma^2 + \omega^2)^2]}
 \nonumber \\
 D_{\rm ind}(\omega) &=&
  \frac{ \gamma^2 \Gamma^2 [ \gamma^2 - (\Gamma^2 + \omega^2)]}{(\Gamma^2 + \omega^2) [\gamma^4 + 2 \gamma^2 (\Gamma^2 - \omega^2) + (\Gamma^2 + \omega^2)^2]} \, .
\end{eqnarray*}
First of all we observe that $D_d(\omega)$ is a positively definite quantity and has no zeros, as one would expect from a conventional transmission coefficient. 
Being a `good' transmission coefficient, its denominator does not have any zeros on the real axis either.
The induced current is finite in almost all parameter regimes. 
However,  for $\gamma^2 > \Gamma^2$ $D_{\rm ind}(\omega)$ changes sign at $\omega=\sqrt{\gamma^2 - \Gamma^2}$. As a result two different generic behaviour types can be observed. For weak $|\gamma| \ll \Gamma$ $I_{\rm ind}(V)$ flows in the direction opposite to $I_{d}(V)$. This regime is largely compatible with the exciton blockade case and we shall call it `exciton' regime. Here the hybridization of the layer states with their `own' electrodes is dominant.
In the opposite case of large $\gamma^2 > \Gamma^2 + V^2$ the currents flow in the same direction and one is confronted with the standard drag situation. These two regimes can also be observed also for $\gamma_0 \neq \gamma_\perp$, see Fig. \ref{fig5}(a). Strong $\gamma$ couplings tend to destroy the excitonic state favoring a rearrangement of individual energy levels in such a way that from the point of view of incoming electrons the constriction is a larger quantum dot rather than two individual layers. Nonetheless, the particle counterflow might be significant. This is the feature our toy model shares with the system considered in the previous Section \ref{s2}. 

In the next step we compute the cross correlation of currents in the layers. The calculations are as straightforward as in the case of the current. The results for the cross cumulant of the induced and direct currents for representative parameter sets are shown in Fig. \ref{fig5} (b). It turns out that the correlations are remarkably strong and as far as their amplitude is concerned their voltage behavior quantitatively follows the results of the incoherent approximation. A fundamental difference is however, the sign of the correlations which never changes sign and is always positive, unlike the situation discussed in Section \ref{s2}.

\section{Summary}
\label{s4}

As we have seen above, a drag counterflow due to the exciton coupling is present in both models.   This is a very remarkable result in view of the fact, that in the first model transport is assumed to be completely incoherent ($\tau_\phi$ is finite and small). On the other hand, while the cross cumulant of the induced and direct currents turns out to be positive in the second model, it is negative in the incoherent approximation. From the physical point of view the fundamental difference between the models is the fully developed gap in the excitation spectrum of the incoherent model. It cannot be generated in the toy model though. Therefore we conclude that for our setups: (i) for the observation of the drag counterflow the gap is not necessary while it is necessary for perfect Coulomb drag; (ii) in the presence of the exciton coupling the counterflow can be achieved even in disordered systems; (iii) current anticorrelations exist only in the gapped systems. We expect these predictions to be observable in the upcoming experiments.\\
The authors would like to thank D. Breyel, S. Maier and F. Dolcini for many interesting discussions. The financial support was provided by CQD of the University of Heidelberg.

\end{document}